\documentstyle[multicol,aps]{revtex}

\def\be {\begin{equation}}
\def\ee {\end{equation}}
\def\ba {\begin{eqnarray}}
\def\ea {\end{eqnarray}}
\def\nn {\nonumber}
\def\a  {\alpha}
\def\b  {\beta}

\def\C  {\Gamma}
\def\C  {\Gamma}
\def\d  {\delta}
\def\D  {\Delta}
\def\e  {\epsilon}

\def\k  {\kappa}
\def\l  {\lambda}

\def\m  {\mu}

\def\O  {\Omega}
\def\p  {\pi}

\def\r  {\rho}

\def\la {\label}
\def\pa {\partial}
\def\f {\frac}
\def\sq {\sqrt}
\def\no {\noindent}
\def\bi {\begin{itemize}}
\def\ei {\end{itemize}}

\def\vs {\vspace}

\def\cals {{\cal S}}
\begin{document}
\draft
\title{Anti-deSitter black holes, perfect fluids, and holography}
\author{Saurya Das and Viqar Husain}
\address{ Dept. of Mathematics and Statistics, }
\address{University of New Brunswick,}
\address{Fredericton, New Brunswick - E3B 5A3, CANADA }
\address{ EMail: saurya,~husain@math.unb.ca   }

\maketitle
\thispagestyle{empty}

\begin{abstract}

We consider asymptotically anti-de Sitter black holes in
$d$-spacetime dimensions in the thermodynamically stable regime.
We show that the Bekenstein-Hawking entropy and its leading order
corrections due to thermal fluctuations are reproduced by a weakly
interacting fluid of bosons and fermions (`dual gas') in
$\Delta=\alpha(d-2)+1$ spacetime dimensions, where the
energy-momentum dispersion relation for the constituents of the
fluid is assumed to be $\epsilon = \kappa p^\alpha$. We examine
implications of this result for entropy bounds and the holographic
hypothesis.

\end{abstract}


\begin{multicols}{2}

\section{Introduction}

It is believed that black holes are thermodynamical objects
with entropy given by one-quarter their horizon areas $A_H$ (in Planck
units) \cite{bh}
\be
S_{BH} = \f{A_H}{4 \ell_{Pl}^{d-2}}~,
\la{bekhaw1}
\ee
where $\ell_{Pl} = (G\hbar/c^3)^{1/(d-2)}$ is the
Planck length and $G_d$ the Newton's constant in
$d$-spacetime dimensions. They satisfy the two laws of
`black hole thermodynamics'
\ba
d(Mc^2) &=& T_H~dS + \mbox{work terms} \\
\D S_{BH} &\geq& 0~
\ea
where $T_H$ is the Hawking temperature. These laws are semi-classical,
and apply to large black holes with  $A_H \gg \ell_{Pl}^{d-2}$.

The idea that black holes are thermodynamic systems in equilibrium
has led to two related areas of study. The first constitutes
suggestions for the underlying microscopic degrees of freedom.
These are postulated based on one's views of what are the
fundamental degrees of freedom arising from a quantum theory of
gravity. The goal is to reproduce the entropy formula
(\ref{bekhaw1}) by tracing a density matrix describing a black
hole over unobserved quantum degrees of freedom. A related
approach involves calculating the number of microstates associated
with macroscopic parameters of a black hole, such as its mass and
charge \cite{vs,ash,anydim}.

The second area of study arises from the observation that the entropy 
of
a confined non-black hole system is proportional to spatial volume, 
rather
than  bounding area. Reconciling this with the fact that black holes 
are
believed to be the most entropic systems leads to a viewpoint known as 
the
holographic hypothesis. This hypothesis states that a $d-$dimensional 
theory
may be encoded exactly in a $(d-1)-$dimensional theory. What remains is 
the
significant task of providing the details of the encoding, the 
``holographic
map.'' A part of this map consists in matching the thermodynamics of
black holes with a suitable system in one lower dimension.

In this paper we focus on the second area of study motivated
by black holes, namely holography. Specifically we address the
following question: {\it What are the criteria under which the
thermodynamics of the AdS-Schwarzschild black hole in spacetime
dimension $d$ is reproduced by a weakly interacting gas in
spacetime dimension $\Delta$?} This question is useful to ask
for at least two reasons: (i) From the viewpoint of ``physical''
holography, one would like to see whether a bulk physical system
in a bounded region (here the black hole) can be described by a
gas in one lower dimension,  and (ii) whether their exist 
thermodynamic
dualities more general than holography, where the difference in
dimension is different from one. We answer the question by showing 
that
the thermodynamics properties of the AdS black hole can be encoded
in a gas of free bosons and fermions, such that the dimension of
the dual gas depends on the dimension of the black hole {\it and}
on the dispersion relation of its constituents. We also show that
the matching of thermodynamics for these two systems
extends to thermal fluctuations.

In the next section, we review recent work on thermal fluctuation
corrections to entropy and its application to black holes. In
section (\ref{aads}) we discuss the thermodynamics of
asymptotically anti-de Sitter black holes and thermal fluctuation
corrections to entropy. In section (\ref{holodual}), we show that
a perfect fluid of bosons and fermions captures the thermodynamics
of these black holes, including thermal fluctuation corrections.
We examine the implications of our results for entropy bounds
\cite{thooft,susskind,bekholo,bousso} in section
(\ref{holography}), and conclude in section (\ref{discussions})
with a summary and list of open questions.

\section{Thermal fluctuations and entropy}

Before we give a systematic derivation of leading order corrections to
entropy of a system due to small thermal fluctuations, we present a
heuristic argument showing how logarithmic corrections to entropy
arise for a black hole.

Consider Wheeler's ``it from bit'' idea, which ascribes one bit of
information (equivalently a spin 1/2) to each Planck area on the
black hole's horizon \cite{thooft,wheeler}. This idea also
incorporates ``holography'' in that degrees of freedom are
associated with area rather than volume. If the area of the black
hole horizon is $A$, the number of microstates is $\Omega = 2^n$,
with $n=A/\ell_{Pl}^{d-2}$, so that entropy $S_{BH} =\ln \Omega$
is proportional to horizon area. This model may be obviously
generalised to associate a spin $s$ with each Planck area without
changing the basic result. The number of microstates is now
$\Omega = (2s+1)^n$, which gives entropy  proportional to area
with a different coefficient.

Corrections to entropy arise in this picture if the component of
total spin along a given axis is fixed. Let $x^i$ of the $n$ spins
have a state $s^i$ chosen from the $(2s+1)$ projections along a
fixed axis. (The total spin component along that axis is then $
\sum_i [x_i (2s/(2s+1))s_i-s ]$.) The number of possible states
${\cal N}$ subject to $\sum_i x_i = n$ is given by the multinomial
distribution function (assuming that the probability for finding
any spin in state $i$ is independent of $i$, and equal to
$1/(2s+1)$ ):
\be
{\cal N} = \f{n!}{x_1! x_2! \cdots x_{2s+1}!}~~.
\la{deg1}
\ee
The entropy $S=\ln {\cal N}$ may be  evaluated using Stirling's 
formula,
assuming $x^i >> 1$ to give
\be
{\cal S}   = a n - s \ln n
= b S_{BH} - s \ln \left( S_{BH} \right)
\la{it1}
\ee
where $S_{BH}= A\ {\rm ln}(2s+1)/l_p^2$ is the entropy with no 
restriction on
total spin component, $a \equiv \sum_i k_i \ln k_i^{-1}$ and 
$b=a/\ln(2s+1)$
are of order unity, and $k_i \equiv x_i/n$. This is a generalisation of 
the
$s=1/2$ case considered in \cite{dmk}. Thus, the leading corrections
are  logarithmic, with the exact coefficient depending on the
spin of the building blocks of the black hole.

We now discuss some generalities of thermodynamic systems in 
equilibrium and
their thermal fluctuations, before proceeding to AdS black holes and
the free boson/fermion gas. This is a review of work 
\cite{bm,rkb,dmb,cap}
which shows that logarithmic corrections to entropy arise from quite 
general
considerations.

The canonical partition function for any thermodynamic system in 
equilibrium
is
\be
Z(\b)
= \int_0^\infty \r(E)~e^{-\b E} dE~~.
\label{part1}
\ee
The density of states $\r(E)$ can be written as an inverse
Laplace transform of the partition function,
\be
\r(E) = \f{1}{2\p i} \int_{-i\infty}^{i\infty}
{Z(\b) e^{\b E}} d\b
= \f{1}{2\p i} \int_{c-i\infty}^{c+i\infty} e^{S(\b)}d\b~~,
\la{density1}
\ee
where

\be
S (\b) =\ln Z + \b E
\la{efn1}
\ee
is the `entropy function'.  A saddle-point approximation of $S(\beta)$
around the equilibrium temperature $\b_0^{-1}$ gives
\be
\r(E) = \f{e^{S_0}}{2\p i} \int_{c-i\infty}^{c+i\infty} e^{1/2
(\b - \b_0)^2 S_0''} d\b ~, \la{saddle} \ee
where $S_0 \equiv S(\b_0)$ and $S_0'' = S_{,\b\b|\b_0}$. By
substituting $\b-\b_0=ix$, and choosing an appropriate contour,
the integral can be evaluated exactly to give
\be
\r(E) = \f{e^{S_0}}{\sqrt{2\p S_0''}} ~~.
\la{corr0}
\ee
The corrected entropy is
\be
{\cal S} := \ln \r(E) = S_0 - \f{1}{2} \ln S_0'' + ~\mbox{(smaller 
terms)}.
\la{entcorr2}
\ee
That the quantity $S_0''$ is indeed a measure of fluctuations of the 
system
can be seen from the  relation
\ba
S''(\b) &=& \f{1}{Z} \left( \f{\pa^2 Z(\b) }{\pa\b^2} \right)
- \f{1}{Z^2} \left( \f{\pa Z}{\pa\b}  \right)^2 \\
&=& <E^2> - <E>^2~,
\ea
where we have used the definitions
\be
<E> = - \left( \f{\pa \ln Z}{\pa \b}\right)_{\b=\b_0}
<E^2> = \f{1}{Z} \left( \f{\pa^2 Z}{\pa \b^2} \right)_{\b=\b_0}~.
\ee
Using the fact that the specific heat of a thermodynamic system in
equilibrium can be written as
\ba
C \equiv \left( \f{\pa E}{\pa T}\right)_{T_0}~
&=& \f{1}{T^2}  \left[ \f{1}{Z} \left( \f{\pa^2 Z}{\pa\b^2} 
\right)_{\b=\b_0}  -
\f{1}{Z^2} \left( \f{\pa Z}{\pa \b} \right)^2_{\b=\b_0} \right] \nn \\
&=& \f{S_0''}{T^2}
\ea
gives
\be
S_0''= C T^2~,
\ee
and hence from (\ref{entcorr2}) that
\be
{\cal S } = S_0 - \f{1}{2} \ln \left( CT^2  \right) + \cdots~~.
\la{master1}
\ee
This formula is applicable to all thermodynamic systems. In particular
it may be applied to black holes by setting $S_0 = S_{BH}$, $T=T_{BH}$ 
and
$C=C_{BH}$ for the specific black hole under consideration.
It is understood that the quantity within the logarithm is divided
by $k_B^2$, the square of the Boltzmann constant.

The application of (\ref{master1}) to black holes was considered
in detail in  \cite{dmb,cap}, where it was shown that it provides
a general approach to understanding corrections to black hole
entropy computed in various models in the literature. For related
and other approaches, see
\cite{fursaev,solo1,other,qg,obregon,carlip,jy,qg1,bss,gks,k,ms,medved,k2,medved2,gg2}.
For other applications of (\ref{master1}), see \cite{appln}.

\section{Asymptotically Anti-de Sitter black holes}
\la{aads}

Here we review thermodynamic properties of the
Ba\~nados-Teitelboim-Zanelli (BTZ) \cite{btz} and
AdS-Schwarzschild black holes, and compute entropy corrections
using the method discussed above.  This method does not apply to
Schwarzschild black holes which have negative specific heat.

For the BTZ black holes
\ba
ds^2 &=& - \left( \f{r^2}{\ell^2} - \f{8 G_3 M}{c^2} \right)~c^2 dt^2  
\nn \\
&+& \left( \f{r^2}{\ell^2} - \f{8 G_3 M}{c^2} \right)^{-1}~dr^2   + r^2 
d\theta^2 ~,
\ea
the entropy, temperature and specific heat are
\ba
S_{BH} &=& \f{2\p r_+ }{4\ell_{Pl}}
= \left( \f{ \ell \p c^2 \sqrt{2}}{\hbar\sqrt{G_3}} \right)
{\sq{M}}  \\
T_H &=& \f{\hbar c r_+}{2\p \ell^2} =
\f{ \hbar \sqrt{2G_3 M}}{\p \ell}
= {\left( \f{\hbar \sqrt{G_3}}{\p \ell c }  \right)^2}~{S_{BH}}
\la{btztemp}  \\
%
C_{BH} &=& \f{dMc^2}{dT_H} = {S_{BH}}~~.
\ea
where $r_+ = \ell\sqrt{8 G_3 M}/c$ is the horizon radius.
Substituting in (\ref{master1}) gives
\ba
{\cal S} &=& S_{BH} - \f12 \ln \left(S_{BH}  S_{BH}^2\right) + \cdots 
\nn \\
&=& S_{BH} - \f{3}{2}\ln \left( S_{BH} \right) + \cdots ~~.
\la{btzcorr}
\ea
This agrees with correction computed using conformal field theory by 
Carlip
\cite{carlip},  including the coefficient $-3/2$ in front of the 
logarithm.

Similarly, for AdS-Schwarzschild black holes in $d$-dimensions,
with a  cosmological constant
$$
\Lambda = - (d-1)(d-2)/2 \ell^2,
$$
the metric is
\ba
&ds^2& = -
 \left(1 - \f{16\p  G_d M}{(d-2)\O_{d-2}c^2 r^{d-3}}
+ \f{r^2}{\ell^2} \right) c^2 dt^2  \nn \\
&+& \left(1 - \f{16\p  G_d M}{(d-2)\O_{d-2}c^2 r^{d-3}}
+ \f{r^2}{\ell^2} \right)^{-1} dr^2 \nn \\
&+& r^2 d\O_{d-2}^2 ~.
\la{adsscmetric}
\ea
The entropy, temperature and specific heat are given by
\ba
S_{BH} &=& \f{ A_{d-2}}{4\ell_{Pl}^{d-2}}
= \f{\O_{d-2} r_+^{d-2}}{4\ell_{Pl}^{d-2}} \la{adsscent}\\
T_H &=& {\hbar c}{}~\f{(d-1) r_+^2 + (d-3)\ell^2}{4\p \ell^2 r_+}
\la{adssctemp} \\
C_{BH} &=& (d-2)\left[ \f{(d-1)r_+^2/\ell^2
+ (d-3)}{(d-1)r_+^2/\ell^2-(d-3)} \right] S_{BH} ~, \la{adsscspht}
\ea
where $\O_{d-2}$ is the area of an unit $S^{d-2}$.
In the so-called `high-temperature limit' $r_+ >> \ell$, a regime in
which the specific heat is positive, the corrected entropy using
({\ref{master1}}) is
\ba
{\cal S} &=&
S_{BH} - \f{1}{2} \ln \left( {S_{BH} S_{BH}^{2/(d-2)}} \right) + \cdots 
\nn \\
&=&  S_{BH} - \f{d}{2(d-2)} \ln \left( {S_{BH}}{} \right) + \cdots 
~~~.
\la{adssccorr}
\ea
It is curious to note that on substituting $d=3$ in the above, the
BTZ result (\ref{btzcorr}) is recovered, although the metric of the 
latter
is not a special case of (\ref{adsscmetric}).

The free energy of the black hole in the limit $r_+>>\ell$ may be 
computed
from Eq.(\ref{adsscmetric}-\ref{adssctemp}):
\be
F_{BH} = Mc^2  - T_{H}S_0 = -c_1 T_H^{d-1}~,
\la{fbh}
\ee
where
\be
c_1 =  \f{\O_{d-2}}{4(d-1) \ell_{Pl}^{d-2}}
\left(\f{4\p \ell^2}{\hbar c (d-1)} \right)^{d-2}~~.
\ee
The entropy may be expressed as
\be
S_{BH} = -\f{\pa F_{BH}} {\pa T_{BH}} = c_1(d-1)T_H^{d-2}.
\la{sbh}
\ee
\section{Dual gas}
\la{holodual}

We now attempt to model the thermodynamic properties of asymptotically
AdS black holes using free bosons and fermions at temperature $T$ in
spacetime dimension  $\D$, and with the dispersion relation
\be
\e = \k p^\a~.
\la{dispersion}
\ee
Although this relation is rather unusual, we will see below that
it leads, via standard statistical mechanics, to the perfect fluid
equation of state
\be
P = \left( {\alpha \over \Delta - 1} \right) \rho,
\label{eos1}
\ee
relating pressure $P$ and energy density $\rho$. This shows, 
independent
of all other results presented here, that the dispersion relation
(\ref{dispersion}) describes the microscopics of the $P=k\rho$ perfect
fluid commonly used  in general relativity. It is interesting that
the dominant energy condition (positive energy density, and  timelike 
or
null energy fluxes), places a restriction on $\alpha$ and $\Delta$:
\be
0 < \alpha \le \Delta - 1.
\label{alphac}
\ee
This shows for example that for $\Delta = 4$, all $\alpha$ such
that $0 < \alpha \le 3$ give physically acceptable fluids. In the
following we derive (\ref{eos1}), and compute other
thermodynamical quantities, for subsequent comparison with black
holes variables.

The thermodynamic potential $\O$ is \cite{landau}
\be
\O = \pm  T \sum_i \ln \left[ 1  \mp e^{(\m-\e_i)/T} \right]
\ee
where $+$ and $-$ signatures refer to bosons and fermions 
respectively,
and $\mu$ is the chemical potential.
In the continuum limit, partial integration gives
\ba
\O &=&  \pm \f{ V_{\D-1} \O_{\D-2} T}{(2\p \hbar)^{\D-1}} 
\int_0^\infty
\ln \left[1 \mp e^{(\m-\e)/T }  \right] p^{\D-2} dp \nn \\
&=&  - \f{ V_{\D-1} \O_{\D-2}}{(\D-1) (2\p \hbar)^{\D-1} 
k^{(\D-1)/\a}}
\times \nn\\
&& \int_0^\infty
\f{\e^{({\D-1})/{\a}}}{e^{({\m-\e})/{T}} \pm 1} d\e~.
\la{omega}
\ea
The energy of the gas is
\ba
E &=& \int_0^\infty \e~d\Gamma (\e)  \nn \\
&=& \f{V_{\D-1} \O_{d-2} }{\a (2\p\hbar)^{\D-1} k^{(\D-1)/\a}}
\int_0^\infty \f{\e^{(\D-1)/\a} d\e}{e^{(\m-\e)/T} \pm 1}~. \ea
Comparing with (\ref{omega}) and using the relation
$\O = -PV_{\D-1}$ gives
\be
E = \left(\f{\D-1}{\a}\right)~P~V_{\D-1}~.
\la{state1}
\ee
This can be written in terms of the energy density
$\r = E/V_{\D-1}$ as
\be
P = k \r~,
\ee
where
\be
k = \f{\a}{\D-1}~.
\label{k}
\ee
This relates spacetime dimension $\D$ and dispersion relation power 
$\a$
to the coefficient $k$ in the equation of state.

The free energy of the gas is
\be
F_{gas} = -  T \ln z = \pm T \sum_{i} \ln (1 \mp e^{-\b \e_i} ) ~.
\ee
The temperature dependence is obtained by partial integration
after taking the continuum limit:
\ba
F_{gas}  &=& \pm \f{ T V_{\D-1} \O_{\D-2}}{(2\p \hbar)^{\D-1}} 
\int_0^\infty
\ln ( 1 \mp e^{-\b \e}) dp~p^{\D-2} ~ \nn \\
&=& -~c_2 V_{\D-1}~T^{\f{\D-1}{\a}+1} ~,
\la{freegas}
\ea
where
$V_{\D-1}$ is the volume of the gas, and
\ba
c_2 &=& \f{\O_{\D-2}}{(\D-1)\k^{(\D-1)/\a} (2\p \hbar)^{(\D-1)}}
\times \nn \\
&~&\zeta \left(\f{\D-1}{\a} + 1 \right)
\C \left( \f{\D-1}{\a} + 1 \right) \times \nn \\
&~& (n_B + n_F  - \f{n_F}{2^{(\D-1)/\a}} )~.
\ea
Here $n_B (n_F)$ is the total number of bosonic (fermionic)
degrees of freedom for the fluid. A similar temperature scaling
for the energy density in four spacetime dimensions was discussed
in \cite{clm}. The entropy is
\be
S_{gas} = - \f{\pa F_{gas}}{\pa T} = c_2
\left(\f{\D-1}{\a}+1 \right) V_{\D-1} T^{\f{\D-1}{\a} }~~.
\la{sgas}
\ee
\section{Matching thermodynamics}

At least two approaches may be taken for comparing
thermodynamics of any two physical systems. A strong duality
might involve equating partition functions, and hence
free energies. This necessarily leads to the matching of
all thermodynamical quantities.

A weaker duality might involve matching only the entropy.
Holographic ideas arising from black hole entropy considerations
suggest only the weaker version. Indeed, the holographic
hypothesis at its basic level is concerned with matching
information, which is kinematical, versus stronger dualities which
involve dynamical comparisons as well.

We follow a weak duality approach, and derive the consequences of
imposing the condition
\be
S_{BH} = S_{gas}~.
\la{sbhgas}
\ee
This weak condition does not automatically lead to matching of
other thermodynamical quantities, such as the specific heat.
Therefore it is still necessary, for example, to compare entropy
corrections due to thermal fluctuations (\ref{master1}), which
depend on temperature and specific heat. If temperature matching
is imposed in addition to (\ref{sbhgas}), it is a stronger
duality,  and leads to entropy corrections matching automatically.
We consider below both the weak and strong cases.

Matching powers of temperature in the entropy formulas
(\ref{sbh}) and (\ref{sgas}) gives our first result
relating spacetime dimensions and $\a$:
\be
\D =\a (d-2) + 1.
\la{match1}
\ee
Thus, given an anti-de Sitter Schwarzschild black hole in
$d$-spacetime dimensions, there is a dual gas in $\D$-spacetime
dimensions which captures thermodynamic information of the black
hole.

Eliminating $\alpha$ in (\ref{alphac})
using (\ref{match1}) gives the relation
\be
 0 < {1\over d-2} \le 1
\ee
for $\Delta\ne 1$, which holds for all $d>2$, and is independent of 
$\Delta$.
Thus (\ref{alphac}) and (\ref{match1}) are consistent. ($\Delta = 1$
is disallowed by (\ref{alphac}) in any case.)

Next we compare the coefficients of the power of temperature
in (\ref{sbhgas}). There are two ways to do this depending on the point 
of
view taken on holography: (i) the holographic degrees of freedom 
associated
with a black hole reside on a surface $r=r_0$ in the black hole 
spacetime, or
(ii)  the holographic degrees of freedom do not reside on any bounding
surface in the black hole background, but rather are defined on
their own flat background spacetime.

According to (i), the temperature $T$ of the gas must be
taken as the red-shifted black hole temperature
\be
 T = \f{T_H}{\sqrt{-g_{00}}} = \f{\ell~T_H}{r_0}
\ee
The black hole entropy in terms of $T$ is
\be
S_{BH} = c_1(d-1)\left( {r_0\over l}\right)^{d-2} T^{d-2}.
\ee
Matching coefficients of powers of $T$ in (\ref{sbhgas}) now gives
\be
c_1 = c_2~\Omega_{\D - 1}~\ell^{d-2}~r_0^{(\a-1)(d-2)}~,
\la{match2}
\ee
where we have used the fact that the gas lives in a subspace of the
surface $r=r_0$ so that its volume is $V_{\D-1} = 
\Omega_{\D-1}r_0^{\D-1}$.

A special case of the relation (\ref{match2}) arises in
the context of the AdS/CFT conjecture. For $d=5$, $\a=1$, $\D=4$ and
$n_B=n_F=8(N^2-1)$, it was shown \cite{gubser,berman}, with
the additional AdS/CFT relation
$$
\p \ell^3/G_5 = 2N^2,
$$
that Eq.(\ref{match2}) is satisfied up to a factor of $4/3$.

According to the point of view (ii), the temperature of the gas is 
equated to the
black hole temperature without any red-shift factors. This leads to the 
following
relation between coefficients:
\be
c_1 = c_2 V_{\D - 1}\ell^{d-2}
\ee
Note that in either case (i) or (ii), the holographic
dimension $\D$ is given by Eq.(\ref{match1}).

From (\ref{match1}), it follows that for the special case $\a=1$,
\be
\D = d -1 ~,
\la{minusone}
\ee
which is normally assumed in the context of holography
\cite{thooft,susskind}. Another indication that $\a=1$ may be
`preferred' lies in the fact that the $r_0$-dependence drops out of
the relation (\ref{match2}) only for this value of $\a$.

For near-extremal stringy black
holes, the near horizon geometry is BTZ \cite{dadg}. Thus, to
describe the thermodynamics of these black holes, one substitutes
$d=3$ in (\ref{minusone}). This gives a $(1+1)-$dimensional gas,
which is known to reproduce the entropy and Hawking radiation
rates of near extremal black holes in string theory
\cite{dma,ag,pmitra}. Some other thermodynamic properties of BTZ
black holes have also been shown to follow form an effective one
dimensional gas \cite{cai}.

We now compare the leading order entropy corrections of the black
hole and gas. Since the correction term is proportional to ${\rm
ln} CT^2$ by Eqn. (\ref{master1}), this comparison is trivial if
both entropy and temperature are matched as for case (ii)
discussed above. However for case (i), the temperatures are not
equated exactly due to the redshift factor associated with the
radial location of the gas. Therefore there are
additional subleading entropy corrections for this case.

Using $C_{gas} = d(F_{gas} + T S_{gas})/dT$, the dimension
matching equation (\ref{match1}) and the perturbation formula
Eq.(\ref{master1}), the corrected entropy of the gas is
\be
\cals^{corr}_{gas}
= S_{gas} - \f{d}{2(d-2)} \ln S_{gas}
- \f{1}{d-2} \ln \left( c_2 V_{\D-1} \right).
\la{gascorr}
\ee

Since $S_{BH} = S_{gas}$,
we see that the leading logarithmic term agrees precisely
with that on the black hole side, Eq.(\ref{adssccorr}),  for all 
spacetime
dimensions, and for any value of $\a$ in the dispersion relation 
(\ref{dispersion}).
The last term in (\ref{gascorr}) depends on the volume of the gas, as 
well the
number of species, and can be interpreted as a finite size effect. 
These
are considered sub-dominant so long as the entropy $S_{BH}$ remains 
large. Thus,
the exact details of the boundary theory are seen to be irrelevant for 
all
for the leading order corrections to match. (Finite size effects were 
also
considered in \cite{finite1}.)

The issue of log corrections for $d=5$ and
$\a=1$ were first analysed in \cite{mukherji}.
However, there the authors had ignored the factor of
$T_H^2$ in (\ref{master1}), by setting the `scale' of the logarithm to
$T_H$ itself. This scale is actually the Boltzmann constant $k_B$,
which has been set to unity here. Consequently, the coefficient in
front of the log term was incorrect there, both for the black hole and
the gas. In addition, the finite-size effect terms were missed.

For $AdS_5/CFT_4$ correspondence, when 
$d=5,~\a=1,~\D=4,~n_b=n_F=8(N^2-1)$, the
coefficient of the log corrections is $5/6$, agreeing perfectly with 
its black
hole counterpart, Eq.(\ref{adssccorr}).

\section{Entropy bounds with AdS black holes}
\la{holography}

In this section we consider the Bekenstein and spherical bounds
for asymptotically anti-de Sitter black holes, and examine the
effects of entropy corrections on these bounds.

The Bekenstein bound states that the entropy of matter in a closed
region of linear dimension $R$ and energy $E$ is bounded
above by the inequality \cite{bekholo,bousso}
\be
S_{matter} \le \f{2\p ER}{\hbar c}~.
\ee

One way to arrive at this result is to consider a `Geroch process,' in
which a matter system in a box is lowered slowly from an asymptotically 
flat
region, and then dropped into the black hole where the box just 
touches
the horizon. The argument assumes that the energy of a floating box
near the horizon is added to the black hole, which increases its 
entropy.
The generalised second law $\Delta S_{Tot} = S^{final}_{BH} - 
(S^{initial}_{BH}
+ S_{matter}) \ge 0$ then leads to the desired inequality. The drop off 
point
occurs where the center of mass of the matter system is a distance $R$ 
from the
horizon, where the energy of the matter acquires a factor 
$\sqrt{-g_{00}}$
relative to its energy at infinity.

This argument may be applied to an AdS-Schwarzschild black hole
background, where the energy of the system is assumed to be $E$ at
a large radius $r>>r_+$.  The energy of the matter a proper
distance $R$ from the horizon is
\be
{\cal E} = \sqrt{-{g_{00}}|_{R}}~E = \left[ \f{(d-1)
r_+}{2\ell^2} \right] ER~. \ee The energy gain of the black hole
is
\be
\d E \le \left[ \f{(d-1) r_+}{2\ell^2} \right] ER~,
\ee
and the corresponding entropy gain is
\be
\d S_{BH} = \f{\pa  S_{BH}}{\pa E}~\d E~.
\la{geroch1}
\ee
Now, in order to find the corrected Bekenstein bound, we use
(\ref{adssccorr}) to compute the entropy derivative. This gives
a `corrected Hawking temperature' $T_H' = \pa \cals/\pa M_{BH}$
given by
\be
T_H' = T_H + \f{d(d-1)\hbar c \ell_{Pl}^{d-2}}{\p (d-2) \O_{d-2} 
\ell^2
r_+^{d-3}}~.
\ee
Substituting this in (\ref{geroch1}) gives
\be
\d S_{BH} = \left[ 1 - \f{2\ell_{Pl}^{d-2}}{\O_{d-2}(d-2)r_+^{d-2}} 
\right]~
\f{2\p ER}{\hbar c}~.
\ee
Imposing the generalised 2nd. law gives
\be
S_{matter} \le  \left[ 1 - \f{2\ell_{Pl}^{d-2}}{\O_{d-2}(d-2)r_+^{d-2}} 
\right]~
\f{2\p ER}{\hbar c}~.
\ee

Notice that the leading term is identical to the Schwarzschild
case \cite{bousso}. The correction,  which depends on the horizon
radius, reduces the bound. The correction term  attains its
maximum when the radius is of the order of Planck length, which
gives

\be S_{matter} \le \left[ 1 - \f{2}{\O_{d-2}(d-2)} \right]~ \f{2\p
ER}{\hbar c}~.
\ee
This equation may be regarded as the modified Bekenstein bound.
(Corrections to the Bekenstein bound from
finite volume corrections were analysed in \cite{kutlar}.)

It is interesting to see what happens to this bound at the
expected saturation point where the system $(E,R)$ is
an AdS-Schwarzschild black hole. The entropy bound becomes
\be
S_{matter}  \le \left( \f{d-2}{2} \right) \left[ 1 -
\f{2}{(d-2)\O_{d-2}} \right] \left( \f{r_+^2}{\ell^2}\right)~
S_{BH}~. \la{bound2}
\ee
Since $r_+/\ell \gg 1$,  the right side is much larger than the black 
hole
entropy. That is, the black hole entropy is far from saturating the 
Bekenstein
bound, implying that the latter is
rather weak in this case. This is a significant departure
from the asymptotically flat case where the upper limit is provided by
the black hole entropy.

The spherical entropy bound similarly undergoes the modification
\be
S_{matter} \le \f{A_H}{4\ell_{Pl}^{d-2}}
- \f{d}{2(d-2)} \ln \f{A_H}{4\ell_{Pl}^{d-2}} ~.
\la{bound3}
\ee
Of course in this case, the black hole saturates this bound, by
definition. From (\ref{bound2}) and (\ref{bound3}) it follows that
the two bounds are {\it not} of the same order for asymptotically
anti-de Sitter black holes, and that the Bekenstein bound is much
weaker than the spherical bound. (After this work was done, we became 
aware
of a recent paper where corrections to Bekenstein and spherical
entropy bounds have been proposed \cite{ggour}. The
correction to the spherical bound proposed in that paper agrees with 
our's,
while that to the Bekenstein bound does not).

Finally we point out the AdS black hole analogues of two related
entropy bounds previously discussed for Schwarzschild
black holes \cite{thooft}. These are bounds on the entropy
of a collection of black holes, and on the entropy of matter
contained in a closed spatial region.

We ask if the spherical entropy bound is satisfied
for a collection of AdS-Schwarzschild black holes with
masses $M_i$ in a bounded region of area $A$. To do so
let us first compare the sum of the entropies
$S_{total}=\Sigma S_i$ of a collection of black holes with
the entropy $S_M$ of a black hole of mass $M=\Sigma M_i$:
\ba
S_{total} &=& \f{\O_{d-2}} {4\ell_{Pl}^{d-2}}
\left[\f{16\p G_d\ell^2}{(d-2) \O_{d-2} c^2} \right]^{(d-2)/(d-1)}~
\nn \\
&& \times  \sum_i M_i^{(d-2)/(d-1)}~,
\ea
and
\ba
S_M &=& \f{\O_{d-2}}{4\ell_{Pl}^{d-2}}
~\left[ \f{16\p G_d\ell^2}{(d-2) \O_{d-2} c^2} \right]^{(d-2)/(d-1)}~
\nn \\
&& \times  \left( \sum_i M_i \right)^{(d-2)/(d-1)}~,
\ea
The relation
\be
\sum_i M_i^{(d-2)/(d-1)} >
\left(  \sum_i M_i\right)^{(d-2)/(d-1)}~
\ee
for powers $(d-2)/(d-1)<1$ gives
\be
S_{total} >  S_M
\ee
This is the opposite of the Schwarzschild black hole result,
and indicates that if $A$ is taken to be the horizon area of
the composite black hole, then $S_{total} > A/4\ell_{Pl}^{d-2}$.
Thus an upper bound on the entropy of a collection of AdS black
holes cannot be derived form this argument.

To obtain a specific example of an entropy bound on a matter
system derived using AdS black holes, we use the gas discussed in
the last section. The energy and entropy of the gas, consisting of
$Z$ species, and with the dispersion relation (\ref{dispersion}),
follows from Eq.(\ref{freegas}), with the replacement $\D
\rightarrow d$  (also see \cite{thooft,bousso,wald}):
\ba
E &=& Z c_2 \left( \f{d-1}{\a} \right) V_{d-1} T^{(d-1)/\a+1} ~, \\
S &=& Z c_2 \left( \f{d-1}{\a} +1 \right) V_{d-1} T^{(d-1)/\a}  ~.
\ea
With $V_{d-1}= \O_{d-1} R^{d-1}$ we can write the entropy
$S(Z,E,R)$ as
\be S = c_3 Z^{\a/(d-1+\a)} R^{(\a(d-1))/(d-1+\a)}
~E^{(d-1)/(d-1+\a)}~, \la{gasentropy}
\ee
where
\be c_3 = c_2^{\a/(d-1+\a)} \O_{d-1}^{\a/(d-1+\a)} \left(
\f{\a}{d-1}  \right)^{(d-1)/(d-1+\a)}~.
\ee
 An entropy bound
arises by requiring that the gas is outside its AdS-Schwarzschild
radius. This is obtained from  the expression for the mass of the
Schwarzschild-AdS black hole in the $r\gg \ell$ limit:
\be
M = \left[ \f{(d-2) \O_{d-2} c^2}{16\p G_d \ell^2} \right]
r_+^{d-1}~.
\ee
As a result the entropy (\ref{gasentropy}) satisfies the inequality:
\be
S < c_4 Z^{\a/(d-1+\a)} \f{A^{(d-1)/(d-2)}}{\ell^{2(d-1)/(d-1+\a)}}
\la{gasbound1}
\ee
where
\be
c_4 = c_3 \left[ \f{(d-2)\O_{d-2} c^4}{16\p G_d}
\right]^{(d-1)/(d-1+\a)} ~~.
\ee
Note that the exponent of $A$ in (\ref{gasbound1}) is greater than
unity, unlike the asymptotically flat case, again resulting in a
weaker bound. Eq.(\ref{gasbound1}) can be written as:
\be
\cals < c_5 \left( \f{\ell_{Pl}}{\l} 
\right)^{\f{(\a-1)(d-1)}{(d-1+\a)}}
\left( \f{\ell_{Pl}}{\ell}  \right)^{ \f{(d-1-\a)}{(d-1+\a)}}~
\f{r_+}{\ell}~S_{BH}~,
\ee
where $S_{BH}$ is the black hole entropy for the same horizon area,
and
\ba
\l^{\a-1} &\equiv& \f{\e \hbar^{\a-1}}{cp^\a} \\
c_5 &=& 4 \a^{\f{d-1}{d-1+\a}} \O_{d-1}^{\f{\a}{d-1}}
\left( \f{d-2}{16\p} \right)^{\f{d-1}{d-1+\a}} \nn \\
&\times&
\left[\zeta\left(\f{d-1}{\a} +1 \right)
\Gamma\left(\f{d-1}{\a} +1 \right) \right]
\nn \\ &\times&
\left[ \left(n_B + n_F - \f{n_F}{2^{(d-1)/\a}}\right)  
\right]^{\f{\a}{(d-1+\a)}}~.
\ea
Given $r_+/\ell \gg1$, the magnitude of the proportionality
factor multiplying $S_{BH}$ on the RHS depend on the two ratios
$\ell_{Pl}/\l$ and $\ell_{Pl}/\ell$.

\section{Discussions}
\la{discussions}

We have shown that the thermodynamics of black holes can be
reproduced by a dual (or holographic) gas with a generalised
dispersion relation. Specifically we have the general result that
for AdS-Schwarzschild black holes in $d$-spacetime dimensions, the
thermodynamics can be encoded in a gas of free bosons and fermions
in $\D = \a (d-2) + 1$ spacetime dimensions. Thus for a given $d$,
a variety of $\D$ can serve our purpose, depending on $\a$. We
have also seen that some results in the AdS/CFT context arise as
special cases of the thermodynamic matching we have used.

We have also derived corrections to entropy bounds, and discussed
specific examples of the bounds, using Schwarzschild-AdS black
holes. An interesting result here is the lack of a spherical bound
for a collection of these black holes.

The dominant energy condition
(\ref{alphac}) and the matching condition (\ref{match1}) do not in
themselves imply that $\Delta < d$, which is an intuitive
expectation from holography. For example, these equations permit
$d=5$ and $\D = 7$ with $\alpha =2$. Thus, a condition over and
above entropy matching is required for this. If $\Delta < d$ is
imposed by hand, it gives the stronger constraint
\be
 \alpha < {d -1 \over d - 2}
\ee
on the coefficient $\alpha$ in the dispersion relation. This is
consistent with both (\ref{alphac}) and (\ref{match1}). Thus it
appears that {\it entropy matching alone does not necessarily
imply dimensional reduction}.

Particles with $\a=1$ do not seem to be necessary in the holographic 
mapping,
although they are sufficient. It is interesting to note 
however,
that for $\a=1$, the
relations that map the entropies  of 
the black hole and gas
become independent of $r_0$, the `location' 
of the dual gas
in the black hole spacetime.
Thus, this may be an 
additional reason to attach a preferred status to
particles satisfying 
the  relativistic dispersion relation.

There are several open questions which would be interesting to
pursue. From the integral in (\ref{saddle}), it is clear that the
next-to-leading-order corrections are difficult, if not impossible
to compute analytically. It may be possible to do this numerically
to find the dependence of the corrections on horizon area. As
noted earlier, the corrected entropy is always less than the
uncorrected one, signifying a reduction in the number of
accessible states, when fluctuations are taken into account. It
may be useful to find an interpretation of this result from the
point of view of information theory, in which decrease of entropy
is associated with an increase in information
\cite{khinchin,shannon}.

An important generalisation of the current
formalism would be to calculate entropy corrections for asymptotically
flat black holes. One way would be to enclose them in
a finite box, such that there is one black hole solution with
positive specific heat \cite{ad}. It would also be interesting
to compare these corrections with those coming from other sources, 
such
as quantum spacetime fluctuations \cite{pm2}.

A related problem concerns  de-Sitter black holes.
Here the temperature and specific heat can be obtained from the
corresponding expressions for asymptotically anti-de Sitter
black holes [Eqs.(\ref{adssctemp}-\ref{adsscspht})],
by substituting $\ell^2 \rightarrow -\ell^2$:
\ba
T_H &=& {\hbar c}{}~\f{- (d-1) r_+^2 + (d-3)\ell^2}{4\p \ell^2
r_+}
\la{dssctemp} \\
C &=& (d-2)\left[ \f{(d-1)r_+^2/\ell^2
- (d-3)}{(d-1)r_+^2/\ell^2-(d-3)} \right] S_{BH} ~. \la{dsscspht}
\ea
Here in the regime in which Hawking temperature is positive,
the specific heat is negative, again signaling an apparent
breakdown of the approach. However, since it has been
claimed that the energy of these black holes
is negative \cite{spradlin,balasub,shanki2}, one may use a modified
definition of specific heat,
namely $C = d(-E)/dT$. This gives a positive $C$.
Corrections to entropy would then be identical to that
for the AdS case, Eq.(\ref{adssccorr}), and it appears that the
mapping of thermodynamics can also be done.

\vs{.4cm}
\no
{\bf Acknowledgements}

The authors would like to thank J. Gegenberg for useful discussions.
S.D. thanks R. K. Bhaduri, P. Majumdar, P. Mitra and
S. Shankaranarayanan for useful correspondence,
and M. M. Akbar, D. S. Berman, M. Cavaglia, A. Dasgupta,
W. R. Knight, R. Maartens and O. Obregon
for helpful comments and discussions. This work was supported in part 
by
the Natural Sciences and Engineering Research Council of Canada.

\end{multicols}

\end{document}